\documentclass[preprint,showpacs,preprintnumbers,amsmath,amssymb]{revtex4}
\usepackage[dvips]{graphicx}

\begin{document}
\title{
A proposal for laser cooling antihydrogen atoms}
\author{P. H. Donnan$^1$}
\author{M. C. Fujiwara$^{2,3}$}
\author{F. Robicheaux$^1$}
\email{robicfj@auburn.edu}
\affiliation{$^1$Department of Physics, Auburn University, AL
36849, USA}
\affiliation{$^2$TRIUMF, 4004 Wesbrook Mall, Vancouver,
British Columbia, V6T 2A3, Canada}
\affiliation{$^3$Department of Physics and
Astronomy, University of Calgary, Calgary, Alberta, T2N 1N4, Canada}
\date{\today}

\begin{abstract}
We present a scheme for laser cooling applicable for an extremely
dilute sample of magnetically trapped antihydrogen
atoms($\bar{\rm H}$). Exploiting and controlling the dynamical
coupling between the $\bar{\rm H}$'s motional degrees of
freedom in a magnetic trap, three-dimensional cooling can
be achieved from Doppler cooling on one dimension using the
$1s_{1/2}-2p_{3/2}$ transition.
The lack of three-dimensional access to
the trapped $\bar{\rm H}$ and the nearly separable nature of
the trapping potential leads to difficulties in 
cooling. Using realistic models for the
spatial variation of the magnetic fields, we find that it should be
possible
to cool the $\bar{\rm H}$'s to $\sim 20$~mK even with these
constraints.
\end{abstract}


\maketitle

\section{Introduction}

Two years ago the ALPHA collaboration demonstrated trapping
of $\bar{\rm H}$ atoms.\cite{ALPHA1} This result was quickly
followed by an improvement in the trapping rate and a
measurement of the lifetime of $\bar{\rm H}$'s in the trap;
$\bar{\rm H}$ were shown to be trapped for longer than 15~minutes.\cite{ALPHA2}
The long time that the $\bar{\rm H}$'s remain trapped is important
because it opens the possibilities to performing measurements
that might require several minutes. ALPHA was able to use
the long trapping time to perform the first measurement of
resonant transitions between $\bar{\rm H}$ bound states.\cite{ALPHA3} 
Recently, the ATRAP collaboration have also published results
where they claim to have trapped $\bar{\rm H}$.\cite{ATRAP}

One of the difficulties in measuring the spectroscopic transitions
in $\bar{\rm H}$ is that the trapped
$\bar{\rm H}$'s are currently at relatively high energies
which will lead to line broadening from the Doppler effect and
from Zeeman shifts.
To date, the trapped $\bar{\rm H}$ are formed using three body
recombination\cite{GO,RH} when
an antiproton ($\bar{\rm p}$) is inside of a positron (e$^+$) plasma. Because three
body recombination is a relatively slow process at the densities
and temperatures in the $\bar{\rm H}$ traps compared to collisional
slowing\cite{HCT,FR}, it is estimated that
the $\bar{\rm p}$ will approximately be in thermal equilibrium with the e$^+$
plasma before recombination occurs. Typical temperatures reported for the
e$^+$ plasma have been a couple 10's of Kelvin. Since the magnetic
trap depth for $\bar{\rm H}$ is only $\sim 1/2$~K, the $\bar{\rm H}$'s
have an energy distribution that extends from 0 to the trap
depth. This was measured in Ref.~\cite{ALPHA2} where the
distribution of $\bar{\rm H}$ annihilations
as a function of time or as a function of axial
position of the trap matched that for the expected $\bar{\rm H}$
energy distribution.

Currently, trapped $\bar{\rm H}$ atoms have energies up to
$\sim 500$~mK, and occupy a large volume of order
(2 cm)$^2\;\times$~30 cm. They are in a strongly non-uniform
magnetic field, varying by order 1~T\cite{ALPHA1}, from the
center to the walls of the trap. While a number of important
experiments have been performed or planned in such a trap,
laser cooling of $\bar{\rm H}$, if achieved, will provide a
major experimental advantage.

Laser cooling will create a cold, and spatially localized
sample of antimatter atoms. Localized atoms will be much less
susceptible to the Zeeman effect, currently a dominant
limitation for microwave spectroscopy\cite{ALPHA3}, and one
of the limitations for future laser spectroscopy\cite{HZ}. 
The lower velocities of cooled atoms will reduce 2nd order
Doppler broadening for 1s-2s two photon spectroscopy.
Importantly, laser cooling will greatly increase the sensitivity
for observing gravitational interaction of antimatter
(see e.g. Ref.~\cite{MCF}).

Only one experiment so far has reported laser cooling of
atomic hydrogen, nearly 20 years ago\cite{SWL}. In that experiment,
$\sim 10^{11}$ atomic hydrogen at 80~mK, pre-cooled via evaporative
cooling, was laser cooled to 8~mK in 15 min. While it is
theoretically possible to cool the $\bar{\rm H}$'s in a similar
manner, $\bar{\rm H}$ laser cooling presents considerable experimental
challenge for several reasons: (1) generation of coherent
radiation at 121.6 nm remains technologically difficult,
due to the lack of convenient lasers and nonlinear crystals
at these wavelengths. (2) experimental requirements for
$\bar{\rm H}$ trapping allow only limited optical access
to the trapped atoms. (3) because of the very low densities
of $\bar{\rm H}$, three-dimensional cooling assisted by
collisionally mixing the degrees of freedom (needed for
laser cooling of atomic hydrogen\cite{SWL}) is prohibitive.
(4) because of the large Zeeman effects, only a small
portion of the trapped atoms resonantly interact with photons. 

Several proposals exist for overcoming some of these
challenges\cite{ZG,DK,WBP}, but none has been experimentally
realized. Reference~\cite{HMP} demonstrated laser cooling
of magnetically trapped Na atoms to $\sim 2$~mK using a
one-dimensional optical molasses. This experiment showed that
it is possible to obtain substantial cooling even with the
severe restriction to one laser.
In this paper, we will investigate whether this simpler
scheme for
laser cooling of $\bar{\rm H}$ will work, and show, via detailed
numerical calculations for ALPHA-type apparatus, that
three-dimension cooling to $\sim 20$~mK should be possible
within realistic experimental and technological constraints. 

In our scheme, the Doppler cooling\cite{MS} will drive
the $1s_{1/2}-2p_{3/2}$ transition with the light being linearly
polarized perpendicular to the quantization axis in order
to drive the $m=1/2$ to $m=3/2$ transition. By driving this
transition, the light scattering does not lead to a
spin-flip which would cause the atom to be ejected from
the trap\cite{SWL}.

While a powerful narrow-line cw Lyman-alpha laser could
eventually offer advantages in laser cooling, development
of such sources remain considerable
challenge\cite{EWH,SKM}. In this work, we consider
the use of a modern pulsed Lyman-alpha source\cite{TM},
whose time-averaged 
(as well as instantaneous) power is much greater than
cw. Because Lyman-alpha generation requires highly non-linear
processes, a pulsed scheme offers overall better cooling
efficiencies, as long as the transition per pulse is not
saturated.

A key feature of the present scheme is the exploitation,
and the control of the dynamical coupling between the $z$-
and $xy$-degrees of freedom. This will allow three dimensional
cooling in an ALPHA-type apparatus in which optical access is
currently limited to one dimension due to constraints
for efficient magnetic trapping as well as high
sensitivity particle detection.\cite{MCF}

While the trapping
fields are clearly non-separable in $x,y,z$ near the walls
of the trap, this is not true near the trap center. The
effective potential energy for the $\bar{\rm H}$'s near the
trap center {\it approximately} has the form $V_1(z) + V_2(r)$
where $r=\sqrt{x^2+y^2}$ and $z$ is along the trap axis. Since
the light is nearly parallel to the trap axis, this nearly
separable potential allows fast cooling of the $z$-motion
but can lead to heating in the $xy$-coordinates.

However, there is some small coupling between the $z$-motion and
the $xy$-motions. This coupling is the conduit through which
we can achieve cooling in all directions. We {\it enhance} this
coupling by the use of non-harmonic magnetic fields in both
$xy$- and $z$- directions, in contrast to standard harmonic magnetic
traps.  In the $xy$-directions, the effective potential is
given by $\sim r^6$, while in the $z$-direction we use total of
five solenoidal coils to produce the nonlinearity. Three
dimensional cooling is possible when the time between photon
scatterings is comparable to or longer than the mixing time
between all of the degrees of freedom. This leads to
non-trivial behavior of the final temperature on the
laser power.

In this paper, we will present results on many of the important aspects
for laser cooling in this constrained geometry. We have investigated
the time dependence of the cooling, the energy distribution
versus detuning, the optimum detuning, etc. In the results
section, we give physical reasons for the difficulties in
trying to laser cool $\bar{\rm H}$.

\section{Numerical Method}

The basic physical situation is that the $\bar{\rm H}$'s classically
move through the trap. When the laser is on {\it and} the
atom is within the waist of the field, the atom can scatter
a photon. The atom receives two momentum kicks for each time a
photon is scattered: when the photon is absorbed the atom
is kicked in the $z$-direction and when the photon is emitted
the atom is kicked in a random direction. The size of each momentum
kick is $h/\lambda\simeq 5.45\times 10^{-27}$kg~m/s. This corresponds
to a velocity kick to an $\bar{\rm H}$ of $\sim 3.3$~m/s. The
$\bar{\rm H}$'s cool when the total momentum kick is opposite the
momentum of the $\bar{\rm H}$. In this section, we describe the
computational techniques we used to model this process.

\subsection{Classical motion}

The $\bar{\rm H}$'s move through a magnetic trap where their
de Broglie wavelength is much smaller than the trap dimensions.
This means we can solve for their motion using classical
forces. The potential energy for the center of mass motion
is equal to $U = -\vec{\mu}\cdot\vec{B}$ where $\mu$ is the magnetic moment of
the $\bar{\rm H}$ in the $1s$ state; $\mu$ is approximately
the magnetic moment of the e$^+$. Since the precession frequency
is much higher than other motional frequency scales, the
angle between $\vec{\mu}$ and $\vec{B}$ is an adiabatic invariant.
This means the orientation of the positron spin with respect
to the magnetic field does not change. Thus, the trapped
$\bar{\rm H}$'s experience a potential $U=\mu B$ where the
$\mu$ is the magnitude of the magnetic moment and $B=|\vec{B}|$.

To compute the force, we need to obtain
\begin{equation}
\vec{F} = -\vec{\nabla}U=-\mu\vec{\nabla}B
\end{equation}
where we need to compute the gradient of the magnitude of the
magnetic field. Because the magnetic field is a very complicated
function of the coordinates, we computed it
numerically using a central two point difference:
\begin{equation}
F_x = -[U(x+dx/2,y,z)-U(x-dx/2,y,z)]/dx
\end{equation}
and similar operations for $F_y$ and $F_z$. We used $dx=dy=dz
=R\times 10^{-5}$ where $R$ is the radius of the trap
$\sim 2.2$~cm. Although this seems a crude approximation,
the error is actually quite small. The error term in the gradient
is $(dx^2/48)U'''$; since $U'''\sim U'/R^2$, this approximation
gives a relative error of $\sim 10^{-11}$ which is comparable
to the round-off error in this approximation.

One of the big problems in the calculation is that we need to
solve for the $\bar{\rm H}$ motion for 100's of seconds. We need to
be careful that there is no energy drift in the calculation
which would either give an unphysical cooling (which would
lead to overly optimistic results) or an unphysical heating
(which would lead to a suppression of the laser cooling).
We found that the adaptive step-size Runge-Kutta algorithm that
worked well for the shorter times needed to model the results
in Refs.~\cite{ALPHA1,ALPHA2,ALPHA3,ALPHA4} was not accurate
enough for the present calculations unless we used very small
time steps. We found that the fourth order symplectic
integrator\cite{EFR,HYo,JCR} worked well for this calculation.
As is usual with symplectic integrators, we found that the
energy varied during the calculation but the variation remained
within a small energy region; thus, there was no energy
drift at long times. We used a time step of 20~$\mu$s
in our calculations.

As with our previous investigations, we launched the $\bar{\rm H}$'s
within an ellipsoidal region with a flat spatial distribution. The
ellipsoid had a scale length of 0.8~mm in the $xy$-coordinates
and scale length of 8~mm in the $z$-coordinate. The initial
velocity distribution was chosen from a thermal distribution
with a temperature $\sim 50$~K except where explicitly stated
otherwise. Since the trap depth is only $\sim 0.5$~K, our
effective velocity distribution is flat in velocity space
within a sphere delimited by the trap depth. Before turning
on the laser pulses, we had the $\bar{\rm H}$'s move through the
trap for 2~s plus a random time between 0 and 0.2~s to model the
fact that there is a delay between the $\bar{\rm H}$ formation
and manipulations done to them. This time delay allows the
$\bar{\rm H}$ to reach somewhat random regions of phase space.

The trapping field is generated using an octupole field to
provide radial confinement and mirror coils to provide
axial confinement.
We used the approximations in Appendix A of Ref.~\cite{ALPHA4}
for the fields from mirror coils and the octupole field.
Instead of two mirror coils, we used 5 coils in order to
mimic the more complicated magnetic geometry in the ALPHA-II
trap. All coils are approximated as two loops with the approximate
vector potential of Eq.~(A.2) of Ref.~\cite{ALPHA4}.
All coils have a radius $a=45.238$~mm
and $\lambda = 0.90230$. The two end mirror coils have a
loop separation of 8.425~mm and a
center-to-center separation of 274~mm. The other 3 coils are equally
spaced between the two end mirror coils. All 3 of these coils
have a loop separation of 8.083~mm.

The purpose of the extra
3 coils is to provide a flatter magnetic field in the center of
the trap. Without the extra coils, the B-field near the center
of the trap is quadratic in $z-z_{mid}$. One coil at the
center can cancel the quadratic dependence and give a B-field
that has a quartic dependence. The coils at the 1/4 and 3/4 position
together with the middle coil can cancel both the quadratic and
quartic dependence. This will give a B-field proportional to
$(z-z_{mid})^6$ near the center of the trap.

We use 3 different currents
through these 5 coils: the end mirror coils have the same current
and the coils at the 1/4 and 3/4 position have the same current.
In most of the calculations, we chose currents to give the flattest
possible B-field. For our coil parameters, this is achieved with
606~A in the end coils, -57.8~A in the coils at the 1/4 and 3/4
position, and -2.5~A in the middle coil. We did perform calculations
with somewhat different currents; these calculations will be
discussed in the Results section.

Figure 1 shows a slice through the magnetic field given in terms
of the Zeeman shift in mK units. The nearly rectangular nature
of the contours highlight the nearly separable nature of the
potential energy experienced by the $\bar{\rm H}$'s at low energy.

\subsection{Light scattering}

To perform realistic calculations, we need to use parameters for
the 121.6~nm light that are within current technical capabilities.
We used parameters suggested to us by T. Momose.\cite{TM}
We assumed a laser with a 10~Hz repetition rate. We assumed the laser pulse
would be on for a short time, $\sim 10$~ns; this time is so short
that we consider both the absorption and re-emission to happen
instantaneously. We used a laser linewidth (FWHM) of 100~MHz.
The total energy in one laser pulse was taken to be 0.1~$\mu$J.
Calculations were mostly performed with the laser propagating exactly along the
$z$-axis and the laser was assumed to be linearly polarized
with a waist radius of 10~mm.
Some calculations were performed at different directions
for laser propagation and will be discussed in the Results section.

We used a semiclassical treatment of laser cooling to model the
interaction between $\bar{\rm H}$'s and the laser field. If the atom
is within the laser waist, the probability
for absorbing a photon during one of the laser pulses is
\begin{equation}\label{eqProb}
P=\frac{3}{8}\frac{c^2}{h f^3}\Gamma_{sp}
\frac{(\Gamma_{sp}+\Gamma_{las})/(2\pi)}{\Delta\omega^2+[(\Gamma_{sp}+\Gamma_{las})/2]^2}
\frac{E_{las}}{\pi w^2}
\end{equation}
where $\Gamma_{sp}$ is the spontaneous decay rate of the $2p$ state
($2\pi 99.7$~MHz $=626$~MHz),
$\Gamma_{las}$ is the laser line width ($2\pi\; 100$~MHz, FWHM), $\Delta\omega$ is
the detuning of the laser (combination of laser detuning, Doppler
shift, and Zeeman shift), $E_{las}$ is the energy in the laser
pulse, and $w$ is the radius of the laser waist. We will measure
the laser detuning from the transition at the minimum
B-field, $B_{min}$, and denote it by $\Delta\omega_0$.
The Doppler
shift is $-\omega_{1s,2p}v_z/c$ where $v_z$ is the z-component
of the $\bar{\rm H}$ velocity and $\omega_{1s,2p}=(E_{2p}-E_{1s})/\hbar$
is the transition frequency between the $1s$ and $2p$ states.
The Zeeman shift is $-\mu_B(B-B_{min})/\hbar$ where $\mu_B$ is the Bohr
magneton because the $m=3/2$ upper state shifts
more strongly in the magnetic field than the $m=1/2$ ground state.
Thus, the detuning in Eq.~(\ref{eqProb}) is given by
\begin{equation}
\Delta\omega =\Delta\omega_0-\omega_{1s,2p}v_z/c-\mu_B(B-B_{min})/\hbar
\end{equation}
where $B$ is evaluated at the position of the $\bar{\rm H}$.

The algorithm to incorporate the photon scattering worked in the
following way. We stepped the $\bar{\rm H}$ using the symplectic
time step of 20~$\mu$s. A laser pulse is sent through the trap
every 5000$^{\rm th}$ step. Nothing happens if the $\bar{\rm H}$ is
outside the waist. If the $\bar{\rm H}$ is within the waist, we
use a random number generator and compare to the probability
to scatter a photon in Eq.~(\ref{eqProb}). If the random number
is smaller than this value, then the atom's velocity is modified
by the two kicks. The two kicks give a change in velocity:
$\vec{v}\rightarrow
\vec{v}+\hat{z}\Delta v+\hat{\nu}\Delta v$ where $\Delta v$ ($\simeq 3.3$~m/s)
is the photon momentum
divided by the $\bar{\rm H}$ mass and the random
direction $\hat{\nu}$ is opposite the direction of the photon
emission. The vector $\hat{\nu}$ is randomly chosen from the photon emission
distribution for a circularly polarized state.

The cooling discussed below should be compared to the cooling in
a three-dimensional (B-field free) optical molasses. The lowest
average atom energy is when the laser detuning is set to
$\Delta\omega_0=-(\Gamma_{sp}+\Gamma_{las})/2$ and gives
an average energy of $E_{av}=3\hbar (\Gamma_{sp}+\Gamma_{las})/4$.
For our laser parameters, the $E_{av}/k_B\simeq 7.2$~mK. The
recoil energy is $E_{rec}/k_B=M\Delta v^2/(2 k_B)\simeq 0.64$~mK.

\section{Results}

There are two important issues that need to be addressed:
what is the best laser detuning and how long is needed to
get substantial cooling. The optimum detuning depends on the
laser power for this system because the laser only directly cools
one direction which is nearly separable from the other
two directions. For our laser parameters, we found
that the detuning should be substantially shifted from the
B-field free optical molasses.
We found optimal cooling with the field free
detuning when we performed calculations where the laser
could cross the trap at a large angle so that there was substantial
components in $z$- and $x$- or $y$-directions, but this will
not be an option for the ALPHA experiment.

We can obtain an overview of the cooling through the time
dependence of the temperature of the trapped $\bar{\rm H}$'s.
Figure~2 shows the average energy of the laser cooled $\bar{\rm H}$'s
as a function of the time that the laser is on. The atoms start
with the distribution that is trapped from a 54~K $\bar{\rm H}$
distribution. The figure shows the results for several possible
detunings of the laser. The solid, dotted, dashed, and dot-dash lines
are for the trapping potential in Fig.~1. Discussed below,
the dash-dot-dot-dot
line is the best cooling that could be obtained when the
trapping potential is only from the two mirror coils at
$\pm 137$~mm.
The value of the detunings are given
in terms of the optimal value of $-(\Gamma_{las}+\Gamma_{sp})/2$
when $B=0$.

This figure shows that the most rapid
cooling occurs at early times and the cooling rate has substantially
slowed near the final times. A pessimistic interpretation is that
this suggests it will take quite a long time to reach the asymptotic
temperature. As will be seen in the figures below, a large fraction
of the $\bar{\rm H}$'s seem to cool to energies under 100~mK within
200~s while some $\bar{\rm H}$'s seem to remain at a few 100~mK. The
hotter atoms have substantially fewer photon scatterings compared
to the colder atoms. Thus, the colder atoms will interact more
with a laser in a spectroscopy experiment which will lead to a
somewhat cooler effective temperature.

We also performed calculations for magnetic fields similar to
the geometry of the original ALPHA experiments. An
example is the dash-dot-dot-dot line in Fig.~2. In this case,
the three middle mirror coils are off and the trapping potential
has a quadratic dependence on the $z$-coordinate. We found the
cooling to be much worse in this case for two reasons. The
quadratic dependence of the magnetic field on $z$ means the
Zeeman shift is substantial for a larger region of space; this
leads to less photon scattering and, hence, a smaller cooling
rate. The other reason is that the quadratic magnetic field
gave even less coupling between the $z$- and $xy$-directions;
this led to a longer mixing time and, hence, a
smaller cooling rate. As with the flat $B$-field case treated
in this paper, we found much improved cooling if the laser is
not constrained to be nearly along the $z$-axis.

Figure~3 shows the time dependence of the energy distribution
for different time windows over a 200 second duration. The
data in Fig.~3 is for when the laser detuning is $4\times$ the
field free optimal value of $-(\Gamma_{las}+\Gamma_{sp})/2$; since we
are using $\Gamma_{las}\simeq\Gamma_{sp}$, this means the
laser detuning is $\simeq -4\Gamma_{sp}$. An
important point is that the distribution is clearly giving
more $\bar{\rm H}$'s at lower energy as the atoms are cooled longer.
This means that substantial laser cooling is possible for this
trap geometry over a time scale where $\bar{\rm H}$'s can be trapped.
The average energy of the $\bar{\rm H}$'s has decreased from
$\simeq 330$~mK to $\simeq 110$~mK over this time. A small fraction,
$\sim 2-3$\%, of the atoms are lost in the cooling process. These
are all atoms that were barely trapped and the first couple of
photon scatterings gave heating instead of cooling due to the
random nature of the laser cooling.

The efficiency of the cooling is naturally of interest. The average
number of photon scatterings by an $\bar{\rm H}$ is $\sim 5$ for each
40~s time window. This gives $\sim 25$ scatterings for the full
200~s of our simulation. There are 2000 laser pulses during this
period which means that there is slightly better than a 1\% chance for
scattering a photon in a laser pulse. For comparison purposes,
the speed of an $\bar{\rm H}$
with 330~mK of kinetic energy is approximately 75~m/s and the
velocity kick from a photon absorption or emission is approximately
3.3~m/s. Thus, each photon scattered provides a substantial amount
of cooling for this detuning and for this duration of cooling.

Figure~4 shows the energy distribution of the $\bar{\rm H}$'s during
the final 40~s time bin as a function of the detuning of the laser
in units of the optimal B-field free detuning. Plots are shown
for detunings of $2\times$, $3\times$, $4\times$, and $5\times$.
All cases started with the same energy distribution of $\bar{\rm H}$'s.
It is clear that the detuning leads to strongly differing
energy distributions. From this figure, it appears that the best
detuning is $3\times$ the field free value because the peak
of the distribution is at the lowest energy. However, the
average final energy is actually lowest for the $4\times$
detuning as seen in the Table I. The average initial energy
is approximately 340~mK for our simulated trap.
\begin{table}
\caption{Average number of scattered photons, $\bar{N}_p$, 
during 200~s cooling and
average final energy $\bar{E}_f$ as a function of laser
detuning, $\Delta\omega_0$ in units of $-(\Gamma_{las}+\Gamma_{sp})/2$.
For this case, the starting distribution was that expected
from the ALPHA experiment.}
\begin{tabular}{c c c}
\tableline
\tableline
\null\hskip 20 pt$\Delta\omega_0$\hskip 20 pt\null&\null\hskip 20 pt$\bar{N}_p$\hskip 20 pt\null&\null\hskip 20 pt$\bar{E}_f$ (mK)\hskip 20 pt\null\\
\tableline
$2\times$& $\sim 50$ &200\\
$3\times$& $\sim 35$ &135\\
$4\times$& $\sim 25$ &115\\
$5\times$& $\sim 15$ &140\\
\tableline
\tableline
\end{tabular}
\end{table}

From Table I, we can gain some insight into how the cooling
process works for this $\bar{\rm H}$ trap. Smaller detuning leads
to more photon scattering but the photon is more likely to
be scattered by atoms with smaller $|v_z|$ because a smaller
Doppler shift brings the photon into resonance. The average
change in energy during a single scattering is
$\Delta E = M(v_z\Delta v + \Delta v^2)$. The $|v_z|$
can be small when the $\bar{\rm H}$ is cold {\it or} when
the atom is moving nearly perpendicular to the $z$-direction.
Thus, the small detuning leads to a lot of scattering without
much energy removed during the scattering event. Since there
is relatively little time between each scattering event, the
$\bar{\rm H}$ does not have sufficient time to mix the motion
in $x,y,z$. But having too large a detuning leads to
a different cooling problem. Simply put, there are too few photons
scattered to give effective cooling during the 200~s simulation
time.

Although Fig.~4 seems to clearly favor the $3\times$ detuning,
Table I gives similar average final energies for $3\times$,
$4\times$, and $5\times$ detuning. This is because the larger
detuning more strongly cools the higher energy part of the
distribution. This suggests that the optimal strategy might
be to change the frequency of the laser so that we
start with large detuning at early times and change the
frequency to smaller detuning
at late times. We found that this did provide more cooling
over the fixed frequency calculation but it was not a
qualitative change. In our calculations, we changed the frequency
linearly with time. For 200~s, the best case we tested started
with $6\times$ detuning initially and finished with $3\times$
detuning. However, the average final energy was 90~mK. Thus,
changing the detuning is probably worth doing only if it is
experimentally easy.

One question we wanted to address is what is the final energy
distribution if one could cool for very long times. To address
this, we started with cold thermal distributions and allowed
them to interact with the laser pulses for 200~s. We found
that atoms with an initial temperature of 50~mK cooled to
$\sim 30$~mK in 200~s with either the $3\times$ or the
$4\times$ detuning. Therefore,
we are presenting results when the atoms start with a
thermal distribution at temperature 30~mK.

Figure~5 is the same as Fig.~3 except for the initial velocity
distribution. In Fig.~5, we start with a thermal distribution
with a temperature of 30~mK ($\bar{E}/k_B=45$~mK)
in order to probe what are the lowest
temperatures that are achievable. For this case,
the energy distribution has settled into its
final value at late times. We find that the energy distribution is well
approximated by a thermal distribution at a temperature of
20~mK ($\bar{E}/k_B\simeq 30$~mK)
which is shown in the inset. We note that the usual
optical molasses temperature would give a temperature
of 4.8~mK for optimal
detuning. Thus, the final temperature for this trap and detuning
is only a factor of $\sim 4$ higher than could be achieved
with a 3-dimensional molasses. We do not have calculations
for the time required to reach the final distribution when
starting from the high temperature case of Fig.~3. However,
our estimates indicate that it should be less than the 1000~s
trapping seen in Ref.~\cite{ALPHA3}.

Figure~6 is the same as Fig.~4 except for the initial velocity
distribution. The $3\times$ and $4\times$ detuning clearly
give better final distributions. Only the $2\times$ detuning
is still evolving at the final time. That case had the distribution
evolving to higher energy at late times.
As in Fig.~4, the $3\times$
detuning has the peak at slightly lower energy than the
$4\times$ detuning. However, the average final energy is
essentially the same for the two detunings.

Table II shows similar data to that in Table I except the
starting energy is a thermal distribution with a temperature
of 30~mK. Except for the
$2\times$ detuning case, the average energy was approximately
constant after 120~s which means Table II gives the asymptotic
average energy for all detunings except the $2\times$ case. Since almost
all of the $\bar{\rm H}$'s are initially trapped, the average initial
energy is 45~mK. The trends present in Table I are reflected in the
data of Table II as well. We think these results are quite encouraging
for laser cooling since the average final energy for a
three dimensional optical molasses is $\sim 15$~mK for the
laser parameters we used in this simulation. The average energy
for the $2\times$ detuning case was still increasing at the final
time; we also found that the average energy for the $2\times$ detuning
was increasing at the final time
when we started with a 50~mK thermal distribution.
The final average energy in that case was $\sim 60$~mK which means
the $2\times$ detuning has an asymptotic average energy which is 
more than two times higher that for $3\times$ or $4\times$ detuning
case.
\begin{table}
\caption{Average number of scattered photons, $\bar{N}_p$, 
during 200~s cooling and
average final energy $\bar{E}_f$ as a function of laser
detuning, $\Delta\omega_0$ in units of $-(\Gamma_{las}+\Gamma_{sp})/2$.
For this case, the starting distribution was thermal at 30~mK.}
\begin{tabular}{c c c}
\tableline
\tableline
\null\hskip 20 pt$\Delta\omega_0$\hskip 20 pt\null&\null\hskip 20 pt$\bar{N}_p$\hskip 20 pt\null&\null\hskip 20 pt$\bar{E}_f$ (mK)\hskip 20 pt\null\\
\tableline
$2\times$& $\sim 85$ &46\\
$3\times$& $\sim 45$ &33\\
$4\times$& $\sim 25$ &32\\
$5\times$& $\sim 15$ &36\\
\tableline
\tableline
\end{tabular}
\end{table}

We have performed calculations for other magnetic field geometries
although we do not present their details. We tried
to increase the cooling rate by increasing the coupling of the
motion in the $z$-direction with the $xy$-directions. We increased
the coupling by deliberately making a small, non-flat potential in the
central region. We observed an increase in the cooling rate
when we increased the current in the central coil
to make a potential hill at the center of the trap. We observed
a {\it decrease} in the cooling rate when we made a potential
dip at the center of the trap by decreasing the current in
the central coil. However, the increase/decrease of the cooling
rate was only apparent
when starting with low energy $\bar{\rm H}$'s because the size
of the perturbations we tested was at the $\sim 10$~mK scale.

\section{Conclusions}

We have performed calculations related to prospects for laser
cooling trapped $\bar{\rm H}$ atoms. Although the standard
methods of laser cooling will work equally well for $\bar{\rm H}$,
the experimental restrictions related to access to the atoms,
the large magnetic fields present in the traps, and the small
wavelength of the light require accurate modeling to address
how much cooling is possible in practice. Our calculations
use accurate magnetic field geometries and realistic laser
parameters. We have found that
an asymptotic temperature only a factor of 2 higher than
for a three-dimensional optical molasses is possible.
We only simulated the case where the laser light was
impinging on the $\bar{\rm H}$'s along the trap axis;
much better cooling is possible
if the laser direction is substantially away from
0$^\circ$ or 90$^\circ$ relative to the trap axis.
Small angles did not have a large effect.

In this paper, we only presented results for nearly flat magnetic fields because
this is clearly the geometry that will be desired for spectroscopic
measurements. If it is possible to use strongly different
$B$-fields and then have them morph to the flat geometry, then
much lower temperatures could be possible. For example, one
might set up a much tighter flat region in $z$ using the
5 mirror coils. After cooling, the $B$-field could be changed
to that in Fig.~1 which would give adiabatic cooling due to
the expansion of the trap region.
Since these possibilities seem likely to be highly machine
dependent, we will save these more complicated situations
for when the experiments are attempted.

This work was made possible in part by a grant of high performance
computing resources and technical support from the Alabama
Supercomputer Authority. This
work was supported by the US National Science Foundation and
the Natural Sciences and Engineering Research Council of Canada,
Nation Research Council Canada/TRIUMF.
PHD supported by the Auburn University Office of the Vice
President for Research through the Undergraduate Research
Fellowship Program.


\begin{figure}[H]
\resizebox{85mm}{!}{\includegraphics{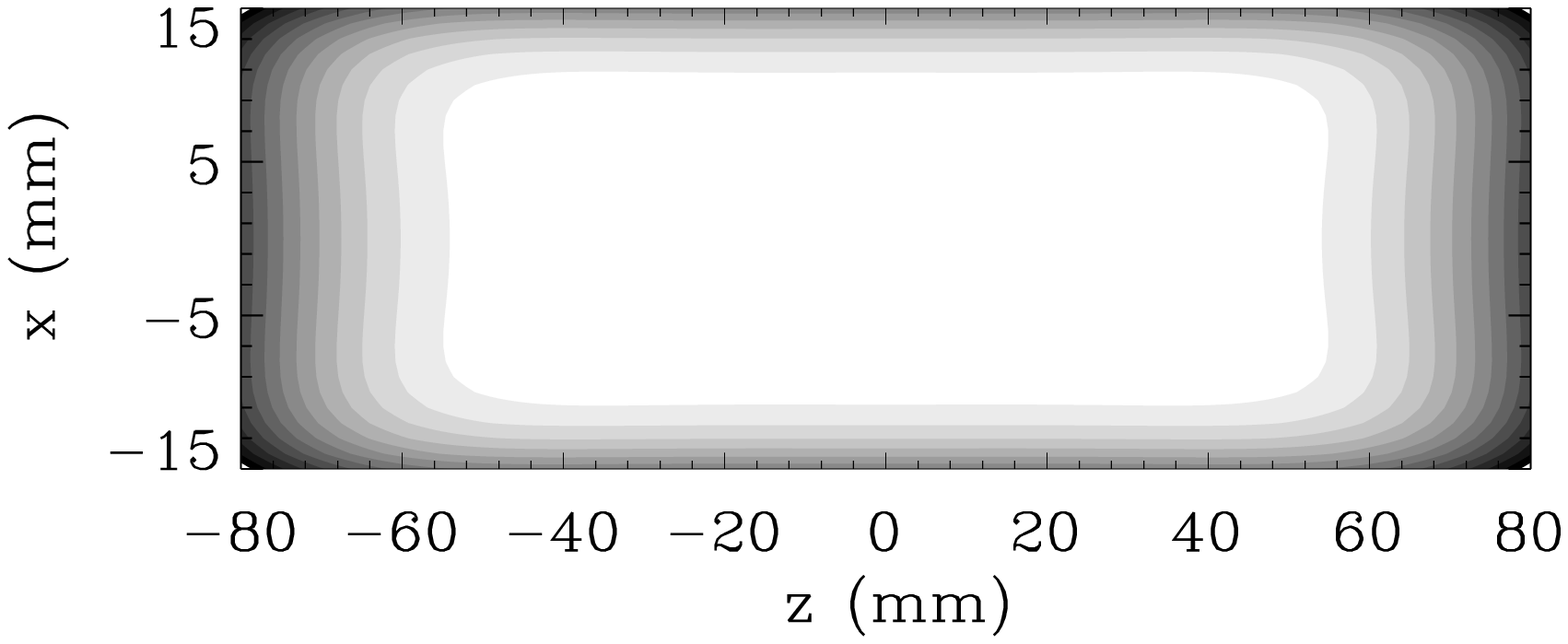}}
\caption{The potential energy experienced by an $\bar{\rm H}$
due to the spatially varying magnetic field. Every contour
represents a change in energy of 10~mK~$\times k_B$.
}
\end{figure}

\begin{figure}[H]
\resizebox{90mm}{!}{\includegraphics{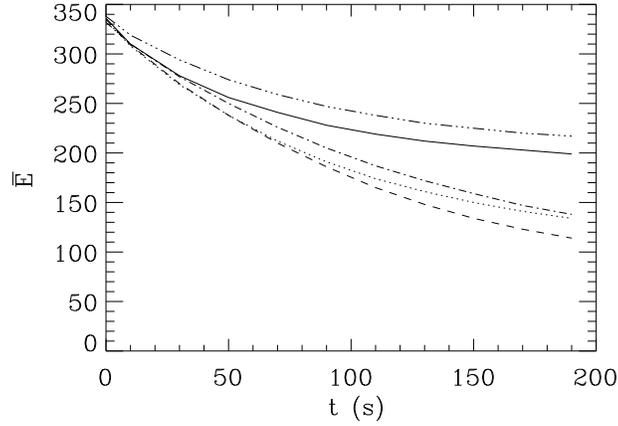}}
\caption{Evolution of the average energy of the $\bar{\rm H}$'s
as a function of the detuning when the magnetic potential
is that shown in Fig.~1. The linetypes are for when
the laser
is detuned $2\times$ the B-field free optimum value (solid line),
$3\times$ detuned (dotted line), $4\times$ detuned (dashed line),
and $5\times$ (dash-dot) line. The dash-dot-dot-dot line is
for the optimum detuning when only the two end magnets are
energized which leads to a trapping potential quadratic in
$z-z_{mid}$.
}
\end{figure}

\begin{figure}[H]
\resizebox{90mm}{!}{\includegraphics{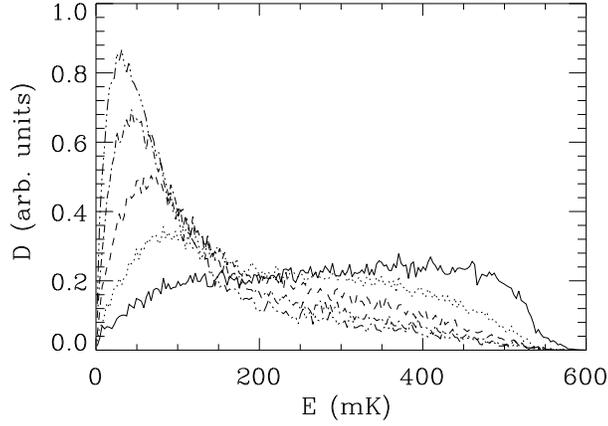}}
\caption{Energy distribution of the $\bar{\rm H}$'s when the laser
is detuned $4\times$ the B-field free optimum value and the
trapping potential is that in Fig.~1. The different
curves correspond to different time windows: 0-40~s is solid,
40-80~s is dotted, 80-120~s is dashed, 120-160~s is dash-dot,
and 160-200~s is dash-dot-dot-dot.
}
\end{figure}

\begin{figure}[H]
\resizebox{90mm}{!}{\includegraphics{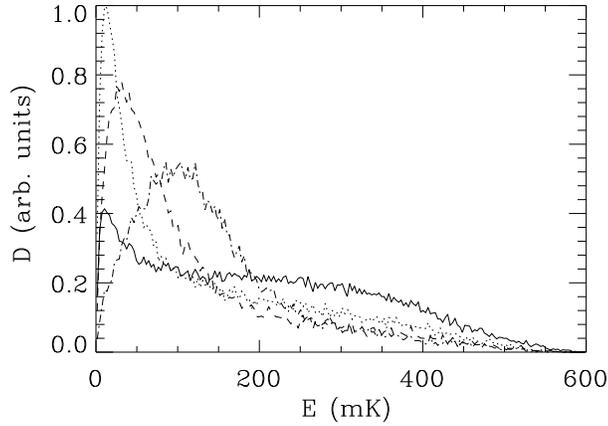}}
\caption{Energy distribution of the $\bar{\rm H}$'s for the time
window of 160-200~s when the laser
is detuned $2\times$ the B-field free optimum value (solid line),
$3\times$ detuned (dotted line), $4\times$ detuned (dashed line),
and $5\times$ (dash-dot) line.
}
\end{figure}

\begin{figure}[H]
\resizebox{90mm}{!}{\includegraphics{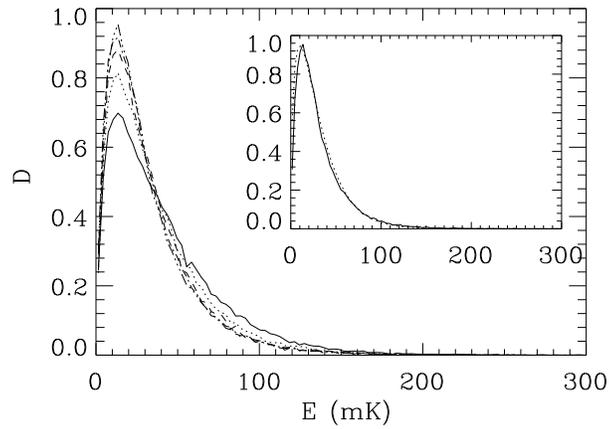}}
\caption{Same as Fig.~3 except that the initial energy distribution
was a thermal distribution at 30~mK (current actual plots are
for 50~mK until the data comes in). The inset shows the energy
distribution in the final window (solid line) and a 20~mK
thermal distribution (dotted line).
}
\end{figure}

\begin{figure}[H]
\resizebox{90mm}{!}{\includegraphics{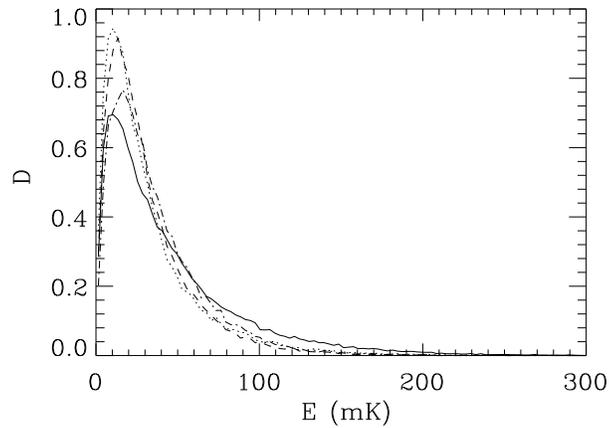}}
\caption{Same as Fig.~4 except that the initial energy distribution
was a thermal distribution at 30~mK.
}
\end{figure}

\end{document}